\documentclass[final,5p,times]{elsarticle}

\usepackage{amssymb}

\usepackage{subfigure}

\usepackage{longtable}
\usepackage{upgreek}

\newcommand{\comment}[1]{}
\newcommand{\nl}{\newline}
\newcommand{\unit}[1]{\ensuremath{\, \textrm{#1}}}
\newcommand{\ave}[1]{\langle#1\rangle}

\journal{Nuclear Physics B}

\begin{document}

\begin{frontmatter}



\title{Neutrino Detection, Position Calibration and Marine Science with
Acoustic Arrays in the Deep Sea}


\author{R.~Lahmann}
\ead{robert.lahmann@physik.uni-erlangen.de}
\address{Erlangen Centre for Astroparticle Physics, Erwin-Rommel-Str.~1, 91058 Erlangen, Germany}

\begin{abstract}
Arrays of acoustic receivers are an integral part of present and
potential future Cherenkov neutrino telescopes in the deep sea. They measure the positions of individual detector elements which vary with time as an effect of undersea currents. At
the same time, the acoustic receivers can be employed for marine
science purposes, in particular for monitoring the ambient noise
environment and the signals emitted by the fauna of the sea. And last but
not least, they can be used for studies towards acoustic detection of
ultra-high-energy neutrinos. Measuring acoustic pressure pulses in
huge underwater acoustic arrays with an instrumented volume of
the order of 100\,km$^3$ is a promising approach for the detection of cosmic neutrinos
with energies exceeding 1\,EeV. Pressure signals are produced by
the particle cascades that evolve when neutrinos interact with nuclei
in water, and can be detected over large distances in the kilometre
range.  In this article, the status of acoustic detection will be
reviewed and plans for the future---most notably in the context of
KM3NeT---will be discussed. The connection between neutrino detection,
position calibration and marine science will be illustrated.

\end{abstract}


\end{frontmatter}


\section{Introduction}

``Of all the forms of radiation known, sound  travels through the sea the 
best''~\cite{urick}.
Hence, sound is 
used in the sea by marine mammals and by humans for purposes of 
communication and positioning.
It is important for astroparticle physics that 
sound waves are furthermore emitted when the local medium heats up 
following the interaction of a neutrino in water. 
As will be elaborated in Sec.~\ref{sec:acou_det_nus}, this effect
allows for the detection of ultra-high-energy neutrinos.
Apart from the design of the detector, the energy threshold is essentially
determined by the
relatively high ambient noise in the sea and by the small
signals expected from neutrino interactions.

The acoustic detection of neutrino reactions 
in principle is possible in any dense homogeneous medium.
In addition to water and ice, 
which are the media of acoustic detection test experiments 
presently or recently conducted,
acoustic detection in salt domes\ \cite{bib:salt-acoustics-arena2005,bib:price-2006} and in 
permafrost\ \cite{bib-permafrost-2008}
has been discussed.
%
In this article only the detection of neutrinos in water will be covered in detail,
in particular in the context of interdisciplinary research and the use
of sound in Cherenkov neutrino telescopes for the position calibration of the optical detectors. 
In Sec.~\ref{sec:test_setups} an overview of current and recent test
setups for the investigation of acoustic neutrino detection techniques
is given
and in Sec.~\ref{sec:acou-bkgr} the acoustic background present in the 
Mediterranean Sea will be discussed.
Monte Carlo simulations required to investigate acoustic detection of neutrinos
are discussed in Sec.\ \ref{sec:MC},
while interdisciplinary use of the data from deep sea acoustic arrays is the subject of Sec.\
\ref{sec:interdisciplinary_coop}. An outlook on the use of acoustics for position calibration and the next step of neutrino detection tests in KM3NeT is given in Sec.~\ref{sec:acoustic} before in 
Sec.\ \ref{sec:summary} conclusions and an outlook are given.

\section{Acoustic Detection of Neutrinos}
\label{sec:acou_det_nus}
Measuring 
pressure pulses in huge underwater acoustic arrays
is a promising approach for the detection of ultra-high-energy (UHE) neutrinos with
energies exceeding 100\,PeV, in particular cosmogenic neutrinos. 
The pressure signals are produced by the
particle showers that evolve when neutrinos interact with nuclei in
water.
The resulting energy deposition in a cylindrical volume of a few
centimetres in radius and several metres in length leads to a local
heating of the medium which is instantaneous with respect to the
hydrodynamic time scales.  This temperature change induces an
expansion or contraction of the medium depending on its volume
expansion coefficient.  According to the thermo-acoustic
model~\cite{Askariyan2,Learned}, the accelerated expansion of the
heated volume---a micro-explosion---forms a pressure pulse of bipolar
shape which propagates in the surrounding medium.
Coherent superposition of the elementary sound waves, produced over the
volume of the energy deposition, leads to a propagation within a flat
disk-like volume (often referred to as {\em pancake})
in the direction perpendicular to the axis of the particle shower.
After propagating several hundreds of metres in sea water, the pulse
has a characteristic frequency spectrum that is expected to peak
around 10\,kHz~\cite{bib:Sim_Acorne,bib:Sim_Acorne2,bib:Bertin_Niess}.
As the attenuation length in sea water in the relevant frequency range
is about
 one to two orders of magnitude larger than that for visible light,
a potential acoustic neutrino detector would require
a less dense instrumentation of a given volume
than an optical neutrino telescope.

\section{Test Setups for Acoustic Neutrino Detection}
\label{sec:test_setups}

Current or recent test setups for acoustic neutrino detection have either been add-ons to 
optical neutrino telescopes or have been using acoustic arrays built for other purposes, 
typically for military use.
%
%
%
In the context of the DUMAND\footnote{
Deep Underwater Muon and Neutrino Detection}
experiment, ideas about adding a large
scale acoustic detector to a deep-sea optical neutrino telescope were
already considered in the 1970s~\cite{bib:roberts-1992}. As the DUMAND
experiment was not realised beyond a prototype phase, acoustic
particle detection was subsequently pursued by the parasitic use of
military arrays. In an early effort starting in 1997 by the 
SADCO\footnote{Sea Acoustic Detector of Cosmic Objects} collaboration,
 a Russian Navy stationary antenna near Kamtchatka
consisting of 2400 hydrophones was used for acoustic particle
detection studies~\cite{bib:sadco-1997}---see also \cite{bib:nahnhauer-arena2010} and references
therein.
                         
\begin{table*}
\begin{tabular}{|l|l|l|l|l|}
\hline
 Experiment & Location & Medium & Sensor Channels & Host  Experiment  \\
\hline
\hline
 SPATS      & South Pole & Ice & 80 & IceCube~\cite{bib:icecube-app:26:155} \\ \hline
 Lake Baikal & Lake Baikal  & Fresh Water & 4  & Baikal Neutrino 
Telescope~\cite{bib:baikal-nt-vlvnt09}    \\ \hline
 O$\upnu$DE        &  Mediterranean Sea  (Sicily)  & Sea Water & 4  &  NEMO\,\cite{bib:nemo-vlvnt09}   \\ \hline
 AMADEUS     &  Mediterranean Sea (Toulon)  & Sea Water & 36  & ANTARES\,\cite{bib:ANTARES-paper}  \\ \hline
 ACoRNE     &   North Sea (Scotland)  & Sea Water & 8  & Rona  military array    \\ \hline
 SAUND    &   Tongue of the Ocean  (Bahamas)  & Sea Water & 7/49$^{(\star)}$  & AUTEC military array     \\ \hline
\hline
\end{tabular}
\caption{Overview of existing and recent acoustic detection test sites.
\nl $^{(\star)}${\footnotesize\,The number of hydrophones was increased from 7 in SAUND-I 
to 49 in SAUND-II, see text.}
}
\label{tab:acoustic_sites}
\end{table*}

%
An overview of experiments in salt water, fresh water and ice that are currently taking data or have done
so until recently is given in Table~\ref{tab:acoustic_sites}.
%
Below, the individual projects will be discussed in some more detail.
\\

\begin{figure}[]
\centering
\includegraphics[width=\columnwidth]{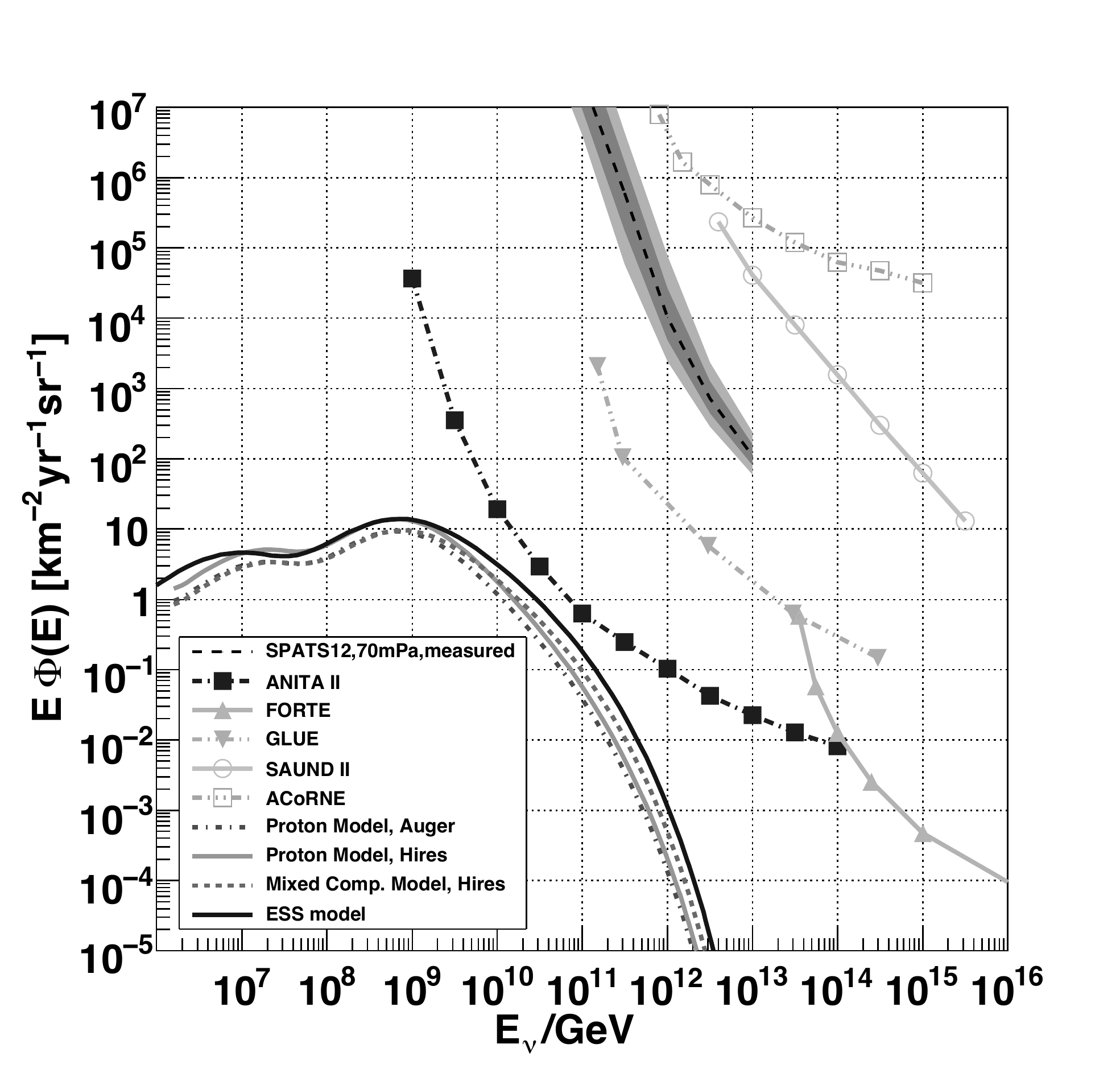}
\caption{
 The neutrino
    flux limit of the 2009 SPATS configuration ($70 \unit{mPa}$
    threshold, $\geq 5$ hits per event) from~\cite{bib:spats-2011-bkgr}.
    The dark grey band ($50$ to
    $100 \unit{mPa}$ threshold) around
    the limit considers uncertainties in absolute noise. The even
    broader light grey band includes additional uncertainties due to
    the choice of different acoustic models.  Experimental limits on
    the flux of ultra-high-energy neutrinos are from ANITA II
    \cite{bib:anita-2010}, FORTE \cite{bib:forte-2004}, GLUE
    \cite{bib:glue-2004}, SAUND II \cite{bib:saund2010}, ACoRNE
    \cite{bib:bevan_ARENA08}. 
Different  models for the cosmogenic flux are shown~\cite{bib:EES-GZKnus-2001,bib:seckel-gzk-www-reduced}.
Figure adapted from~\cite{bib:spats-2011-bkgr}.
}
\label{limits_spats2011} 
\end{figure}

The {\bf SPATS (South Pole Acoustic Test Setup)} 
project~\cite{bib:karg-arena2010,bib:SPATS-design-perform-2011}, 
deployed up to a depth of 500\,m in the upper part of four boreholes of the
IceCube Neutrino Observatory, has continuously monitored the noise in 
Antarctic ice at the geographic South Pole since January 2007.
As acoustic properties, in particular the absorption length and the speed
of sound, have been subject to fewer experimental studies 
for ice than for water, these properties have been instigated with 
SPATS~\cite{bib:spats-2010-speed,bib:spats-2011-atten}. 
Based on 8 months of observation,
a limit on the neutrino flux above $10^{11}$\,GeV has been 
derived~\cite{bib:spats-2011-bkgr}, see Fig.~\ref{limits_spats2011}.
\\

In {\bf Lake Baikal}, an antenna consisting of four hydrophones in a 
tetrahedral arrangement with equal interspacings of the hydrophones of 
$1.5\,\mathrm{m}$
has been placed at 150\,m depth~\cite{bib:baikal}.
Fresh water has the advantage over sea water in that the attenuation
length is roughly one order of magnitude larger in the frequency range
of 10 to 100\,kHz. 
However, conditions in Lake Baikal
are not particularly favourable for acoustic neutrino detection, since in the 
deep zone of the lake the water temperature is only 
$1.5-2^\circ\mathrm{C}$ higher than
the maximum density at the respective 
depth~\cite{bib:baikal-arena2010,bib:baikal-acoust-icrc09}.
The thermal expansion coefficient hence is close to zero and the
Gr\"uneisen parameter small. 
The observed noise level depends mostly on surface conditions and in
the frequency range of 5 to 20\,kHz has a value of a few mPa.
\\

The {\bf O$\upnu$DE (Ocean noise Detection Experiment)} project
at the site of the NEMO\footnote{Neutrino Mediterranean Observatory}
Cherenkov neutrino detector~\cite{bib:nemo-vlvnt09}
has performed long term noise studies at 2050\,m depth, 
25\,km east of Catania (Sicily)
in the Mediterranean Sea at the location 37$^\circ$30.008'N, 15$^\circ$23.004'E.
Phase I operated from January 2005 until November 2006.
It employed 4 hydrophones forming a tetrahedral antenna with side lengths
of about 1\,m. 
In an analysis carried out with data recorded during 13 months between
May 2005 and November 2006~\cite{bib:noise_ONDE}, 
the
average acoustic sea noise in the band 20 to 43\,kHz was measured as 
$5.4\pm2.2\,\mathrm{(stat)}\pm0.3\,\mathrm{(sys)~mPa\,(RMS)} $.
In 2011, the deployment of a new hydrophone antenna is planned in the
context of the NEMO-II project.
\\


The {\bf SAUND (Study of Acoustic Ultra-high energy Neutrino Detection)}
experiment~\cite{bib:Lehtinen-2002} 
employed a large hydrophone
array in the U.S. Navy Atlantic Undersea Test
And Evaluation Center (AUTEC)~\cite{bib:Lehtinen-2002}. 
The array is located in the Tongue of the Ocean, a deep tract of sea
in the Bahama islands at approximately 24$^\circ$30'N and
77$^\circ$40'W.
In the first phase SAUND-I, 7 hydrophones arranged over an area of  
$\sim$250 km$^2$ were used for studies of UHE neutrino detection.
The hydrophones were mounted on 4.5 m booms standing vertically
on the ocean floor at about 1600\,m depth. The horizontal spacing between
central and peripheral hydrophones was between 1.50
and 1.57 km.
After the upgrade of the array, 49 hydrophones that were mounted 5.2 m
above the ocean floor, at depths between 1340 and 1880\,m were
available.  The upgraded array spans an area of about 20 km $\times$ 50 km
with spacing of 3 to 5 km.  This array was used in the second phase
SAUND-II. Neutrino flux limits were derived with SAUND-I for a lifetime of 195
days~\cite{bib:saund2004} and for SAUND-II for an integrated lifetime
of 130 days~\cite{bib:saund2010}, see Fig.~\ref{limits_spats2011}.
\\

The {\bf ACoRNE (Acoustic  Cosmic Ray Neutrino Experiment)}
project~\cite{bib:acorne} utilises the Rona hydrophone
array, situated near the island of Rona between the Isle of Skye and
the Scottish mainland. At the location of the array, the sea is about
230\,m deep.
The ACoRNE Experiment uses 8 hydrophones, anchored to the sea bed and
spread out over a distance of about 1.5\,km. Six of these hydrophones
are approximately in mid-water, one is on the sea bed while the last
one is about 30\,m above the sea bed.  The ACoRNE collaboration has
derived a flux limit on UHE neutrinos~\cite{bib:bevan_ARENA08} which
is shown in Fig.~\ref{limits_spats2011}. 
\\

\begin{figure}[t]
\centering
\includegraphics[width=\columnwidth]{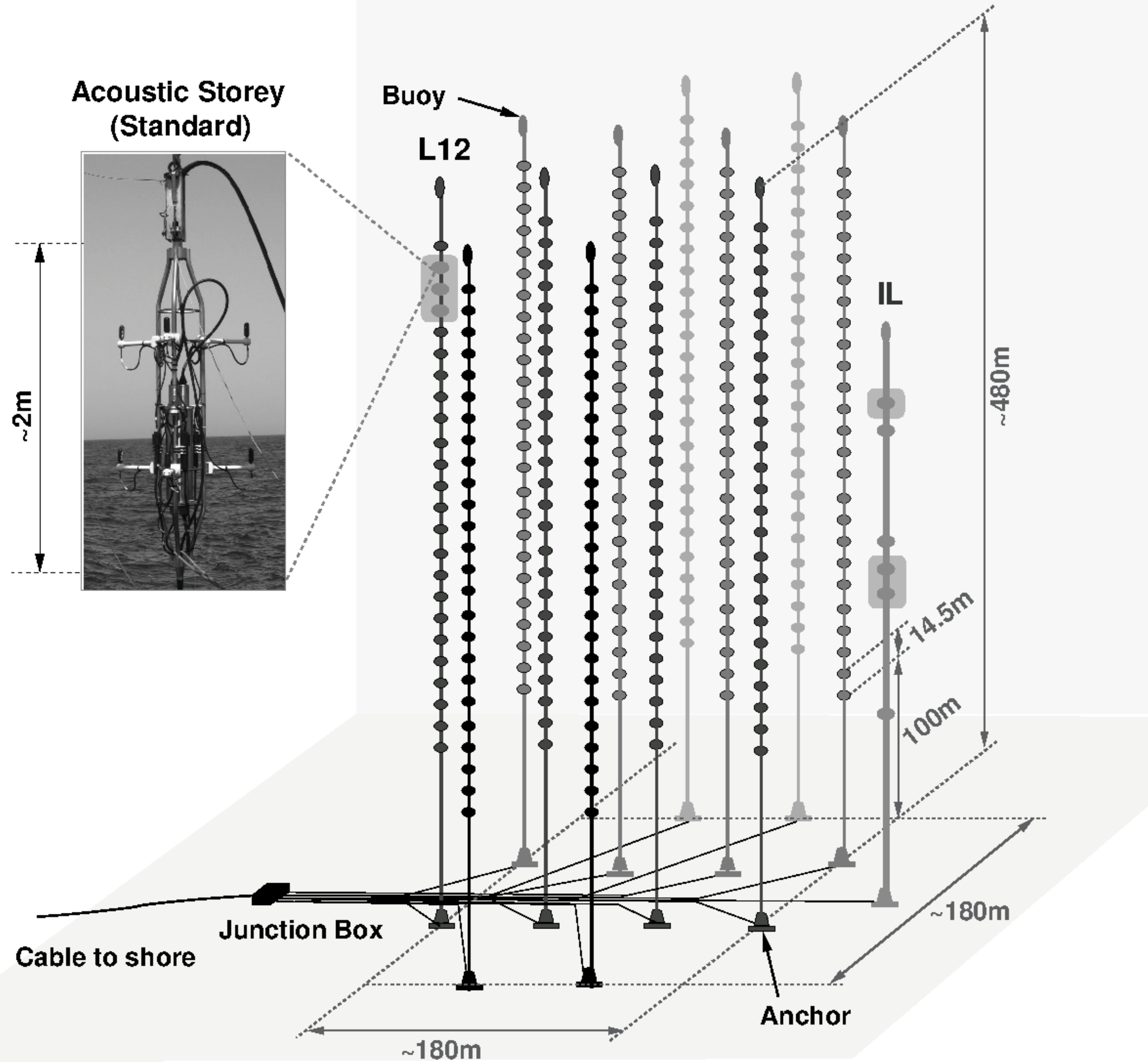}
\caption{A sketch of the ANTARES detector. 
The six acoustic storeys are highlighted and a photograph of a storey in
standard configuration is shown. 
L12 and IL denote the 12th detection line and the Instrumentation Line, 
respectively.}
\label{fig:amadeus_schematic} 
\end{figure}

The {\bf AMADEUS (ANTARES Modules for the Acoustic Detection Under the Sea)} 
project%
~\cite{bib:amadeus-2010} 
was conceived to perform a feasibility study for a
potential future large-scale acoustic neutrino detector
in the Mediterranean Sea. For this purpose, 
a dedicated array of acoustic sensors was integrated into the
ANTARES\footnote{
Astronomy with a Neutrino Telescope and Abyss environmental Research} neutrino telescope~\cite{bib:ANTARES-paper}. 
%
A sketch of the detector, with the AMADEUS modules highlighted, is
shown in Figure~\ref{fig:amadeus_schematic}.  The detector
is located in the Mediterranean Sea at a water depth of about 2500\,m,
roughly 40\,km south of the town of Toulon at the French coast at the
geographic position of 42$^\circ$48$'$\,N, 6$^\circ$10$'$\,E.  ANTARES was
completed in May 2008 and comprises 12 vertical structures, the {\em
  detection lines}.  
Each detection line holds up to 25 {\em storeys}
that are arranged at equal distances of 14.5\,m along the line.
A standard storey 
holds three {\em Optical Modules},
each one consisting of a photomultiplier tube inside a
water-tight pressure-resistant glass sphere.
A 13th line, called the {\em Instrumentation Line (IL)}, is equipped with
instruments for monitoring the environment. It holds six storeys.

Within the
AMADEUS system~\cite{bib:amadeus-2010}, acoustic sensing is integrated 
in the form of {\em
  acoustic storeys} that are modified versions of standard ANTARES
storeys, in which the Optical Modules are replaced by custom-designed
acoustic sensors.  
Dedicated electronics is used for the amplification, digitisation
and pre-processing of the analogue signals.  
Six acoustic sensors per storey were implemented, arranged at
distances of roughly 1\,m from each other. 
%
The AMADEUS system comprises a total of six acoustic storeys: three on
the IL 
and three on the
12th detection line (Line 12).
%
In the following, results from the AMADEUS device will be presented.

As the upper part of each line is not fixed but instead moves with undersea currents,
ANTARES also contains a dedicated acoustic positioning system to continuously monitor the positions of the Optical Modules. Efforts for KM3NeT are going in the direction of
designing one common acoustic system for both applications, as will be discussed in Sec.~\ref{sec:km3net}.

\section{Acoustic Background in the Mediterranean Sea}
\label{sec:acou-bkgr}
\subsection{Ambient Background}

Ambient noise, which can be described by
its characteristic power spectral density (PSD), is caused
by environmental processes and 
determines the minimum
pulse height that can be measured, if a given signal-to-noise ratio (SNR)
can be achieved with a search algorithm. 
To measure the ambient background at the ANTARES site, data from one
sensor on the IL taken from the beginning of 2008 until the end of
2010 were evaluated.
After quality cuts, 27905 minimum bias samples (79.9\% of the total
number recorded in that period) remained for evaluation, each
sample containing data continuously recorded over a time-span of
$\sim$10\,s.
For each of these samples, the noise PSD
(units of $\mathrm{V^2/Hz}$)  was integrated in the
frequency range $f = 10 - 50$\,kHz, 
yielding the square of the ambient noise
for that sample, as quantified by the output voltage of the hydrophone.  
Preliminary studies using the
shower parameterisation and algorithms from~\cite{bib:Sim_Acorne2}
indicate that this range optimises the SNR for
the expected neutrino signals.

\begin{figure}[ht]
\centering
\includegraphics[width=8.0cm]{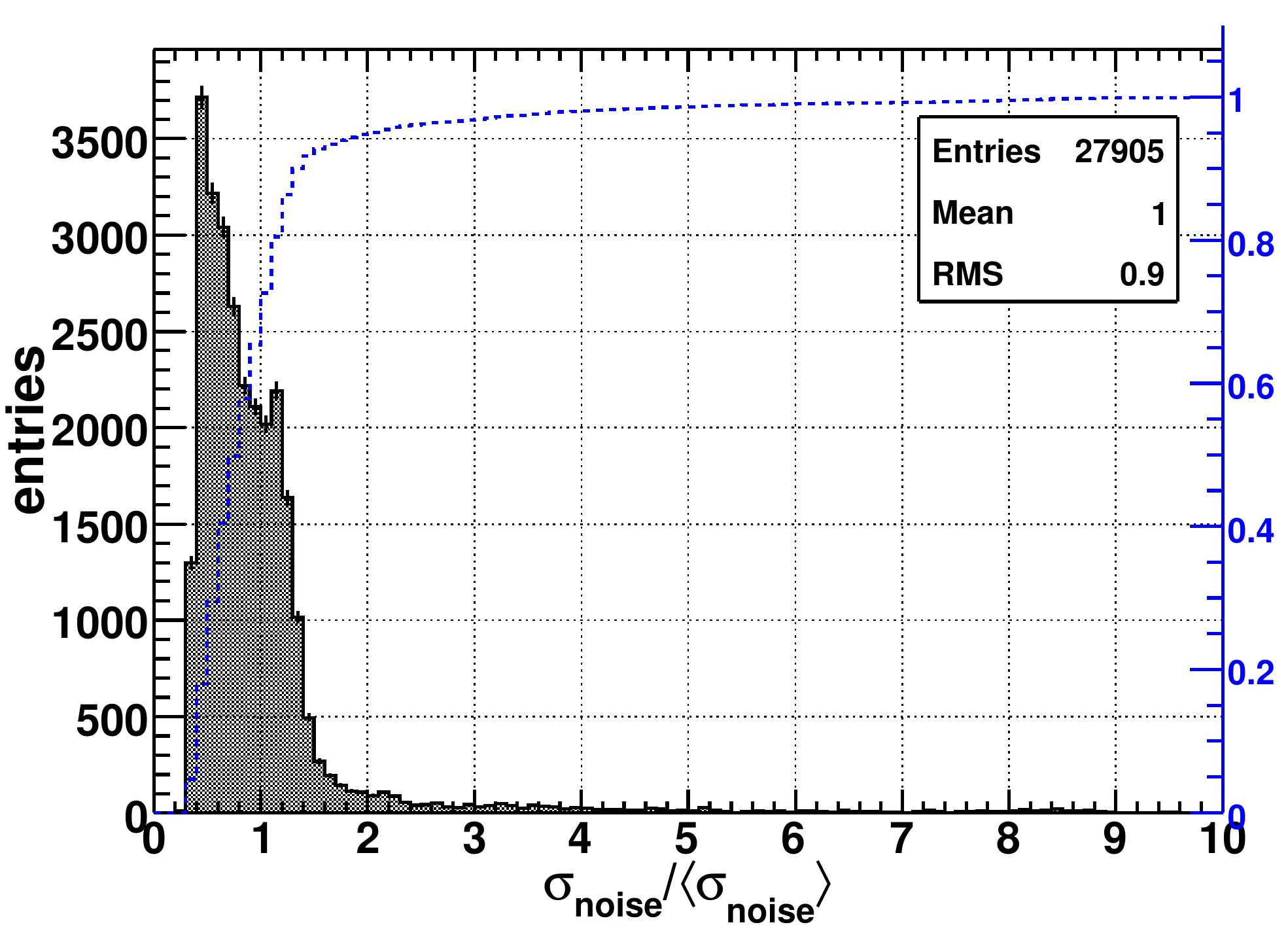}
\caption{
Frequency-of-occurrence distribution
for the ambient noise in the range $10 - 50$\,kHz, relative to the mean ambient 
noise   recorded over the 
complete period of three years which was used for the analysis
(left scale, filled histogram).
Also shown is the cumulative distribution, normalised to
the total number of entries of the distribution (right scale, dotted
line). 
}
\label{fig:noise_distr}
\end{figure}

The frequency-of-occurrence distribution of the resulting noise
values, relative to the mean noise over all samples, is shown in
Fig.~\ref{fig:noise_distr}.  Also shown is the corresponding
cumulative distribution.  For 95\% of the samples, the noise level is
below $2\ave{\sigma_\mathrm{noise}}$, demonstrating that the ambient
noise conditions are stable.

All sensors have been calibrated in the laboratory prior to
deployment.  The absolute noise level can be estimated by assuming a
constant sensor sensitivity\footnote{The ambient noise is originates
  mainly from the sea surface and hence displays a directivity which
  has to be folded with the variations of the sensitivity over the
  polar angle to obtain an effective average sensitivity.  For the
  results presented here, the noise has been assumed to be isotropic.
} of $-145\pm2$\,dB\,re\,1V/$\upmu$Pa.  With this value, the mean
noise level is $\ave{\sigma_\mathrm{noise}} = 10.1^{+3}_{-2}$\,mPa
with the median of the distribution at $8.1$\,mPa.

Currently, the detection threshold for bipolar signals corresponds to
a SNR of about 2 for an individual hydrophone.  For this SNR, the
median of the noise distribution corresponds to a signal amplitude of
$\sim$15\,mPa, equivalent to a neutrino energy of $\sim$1.5\,EeV at a
distance of 200\,m~\cite{bib:Sim_Acorne}.  By applying pattern
recognition methods that are more closely tuned to the expected
neutrino signal, this threshold is expected to be further reduced.

\subsection{Transient Sources}
Transient sources, e.g.\ from sea mammals and shipping traffic, may create signals containing 
the characteristic bipolar pulse shape that is expected from 
neutrino-induced showers. 
In order to reduce the data volume,
the AMADEUS system employs an online
pulse-shape-recognition trigger which is sensitive to the bipolar pulse expected from neutrino
interactions. This trigger
selects events with a wide range of shapes. 
For further offline data reduction,
a classification scheme is being developed
which allocates triggered events to one of four classes: 
genuine bipolar events that are compatible
with signals expected from neutrinos (``neutrino-like events''), 
multipolar events, reflections of
signals from the acoustic emitters of the ANTARES positioning system, and random 
events, where the latter class
contains all events that do not fit into any of the other classes.
For the classification, simulated signals representing the four
classes in equal proportions were produced and a set of features
extracted which 
are highly discriminatory between the classes.
This feature vector is then
fed into a machine-learning algorithm~\cite{bib:neff_arena2010}.
Classification is performed for the signals from individual hydrophones.
Subsequently,
the results from individual hydrophones are combined to derive a 
classification for a given acoustic storey.
Several algorithms were investigated, the best of which yielded a failure rate (i.e.\ wrong decision
w.r.t. simulation truth) at the 1\%-level when applied to the two
signal classes ``neutrino-like'' and ``not neutrino-like''.
After selecting 
neutrino candidates on the level of a storey, measurements from
multiple storeys can be combined to search for patterns that are
compatible with the characteristic ``pancake'' pressure field resulting
from a neutrino interaction.

\comment{ 
\subsubsection{Source Position Reconstruction}
\label{sec:source_dir_reco}
The sensors within a cluster allow for efficient triggering of
transient signals and for direction reconstruction.  The combination
of the direction information from different acoustic storeys yields
the position of an acoustic source.
\begin{figure}[ht]
\centering
\includegraphics[width=9.0cm]{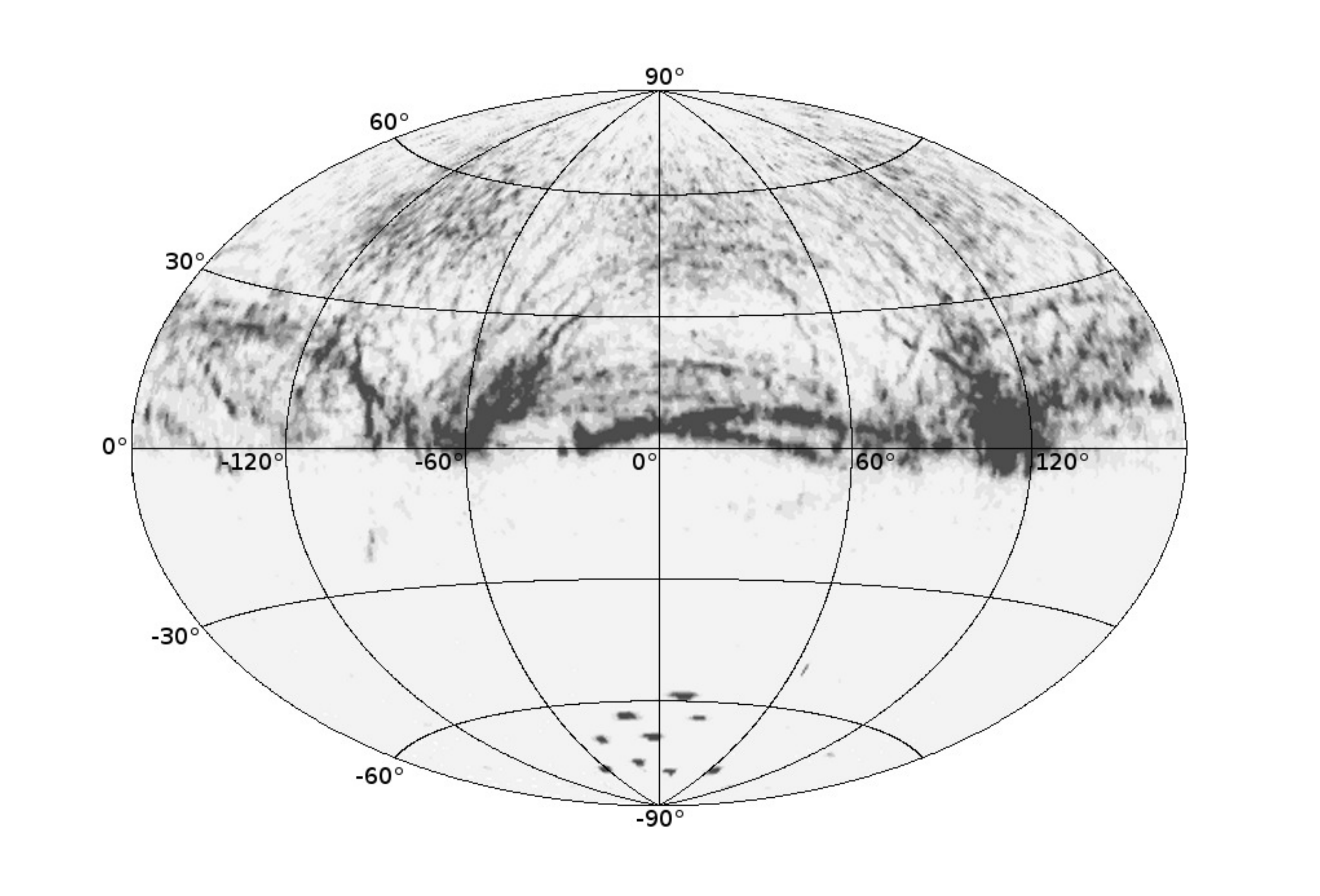}
\vspace{-7mm}
\caption{
Map of directions of sources as reconstructed with an acoustic storey on
Line 12. Zero degrees in azimuth correspond to the north direction, the 
polar angle of zero corresponds to the horizon of an observer on the
acoustic storey. At the bottom, the signals of the emitters
of the ANTARES positioning system are visible. 
}
\label{fig:skymap}
\end{figure}
Figure~\ref{fig:skymap} shows the reconstructed directions of all
sources that were triggered during a period of one month. 
The dark bands of increased acoustic activity 
can be associated with shipping routes and  points of
high activity with the directions of local sea ports. 
It is obvious from Fig.~\ref{fig:skymap} that a fiducial volume for the 
determination of the background rate of bipolar events must exclude the 
sea surface.
} 

\section{Monte Carlo Simulations}
\label{sec:MC}
\begin{figure}[tbh]
\centering
\includegraphics[width=8.4cm]{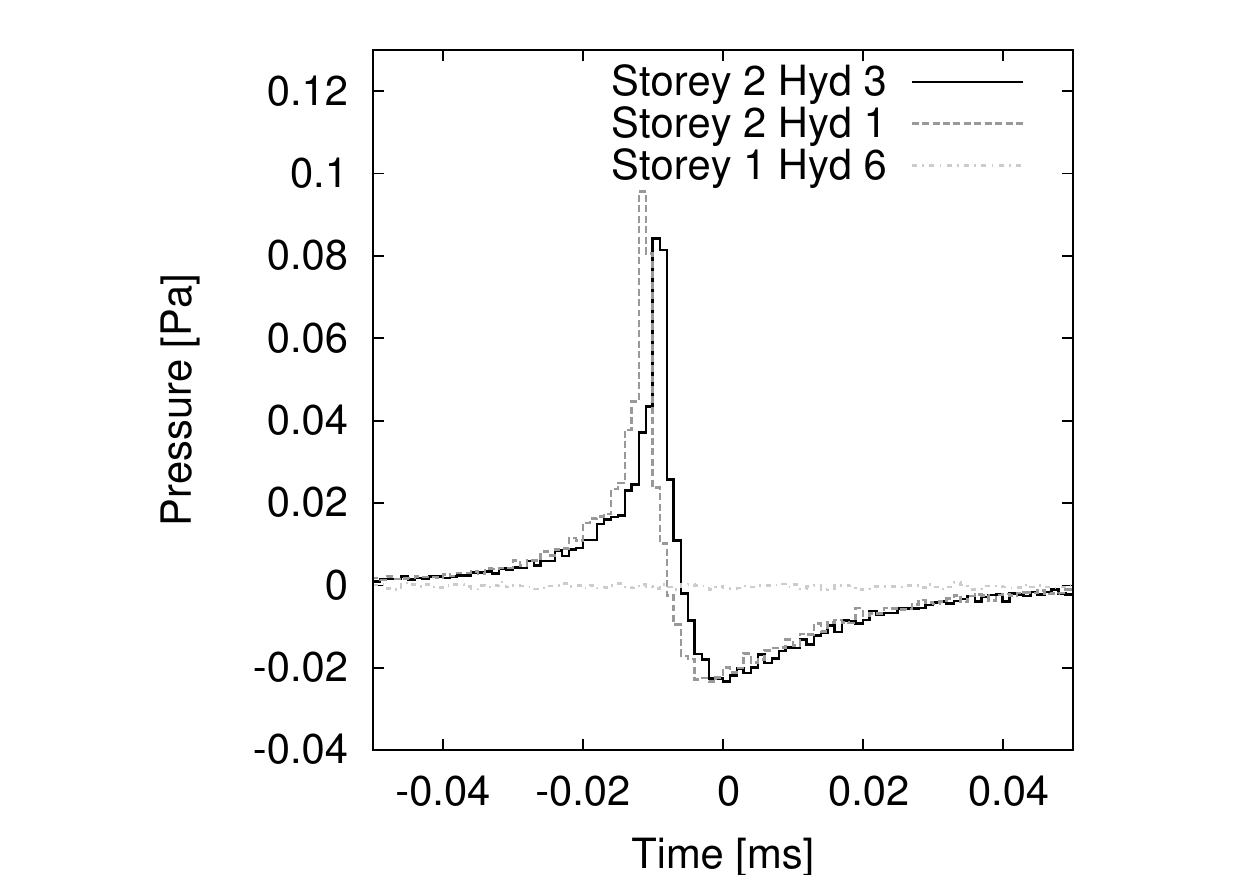}
\caption{Simulated acoustic signals as recorded with hydrophones in
  two acoustic storeys with a vertical spacing of 14.5\,m. See text
  for details. For Storey 2, signals from two different hydrophones are
  shown.  }
\label{fig:bip}
\end{figure}

Monte Carlo simulations based on
~\cite{bib:Sim_Acorne,bib:Sim_Acorne2} are currently 
being implemented
for the AMADEUS detector setup.
%
%
The pressure pulse  received by a hydrophone resulting from the energy deposition of a
$10^{10}$\,GeV  is shown in Fig.~\ref{fig:bip}. The corresponding
neutrino interaction was generated such that the centre of the
hadronic shower for a vertically downgoing neutrino lies within the
same horizontal plane as the storey denoted ``Storey 2'', at a distance
of 200\,m.
This way, the storey lies within the ``pancake'' of the pressure
field.  On a storey 14.5\,m below that storey, denoted ``Storey~1'',
no signal is observed.  This configuration corresponds to two adjacent
acoustic storeys on L12 or the two lowermost storeys on the IL, see
Fig.~\ref{fig:amadeus_schematic}.  This simulation
illustrates the characteristic three-dimensional pattern expected from
neutrino-generated pressure waves. A more detailed discussion can be found in
\cite{bib:neff-vlvnt2011}.

\section{Interdisciplinary Cooperation}
\label{sec:interdisciplinary_coop}
Signals from marine mammals and other environmental sour\-ces
constitute background  for the
acoustic detection of neutrinos. In particular, dolphins emit whistles with
frequency spectra that resemble those expected from neutrino interactions.
Dolphins can dive to depths of $\sim 500\,\mathrm{m}$ and their whistles
constitute the main background of transient
signals recorded by the AMADEUS pulse shape recognition trigger.
On the other hand, these signals
are of great interest to environmental and marine science and
the acoustic monitoring of the deep sea has
a large potential for interdisciplinary research.
As a consequence, efforts for
acoustic detection of neutrinos are pursued in cooperation with marine
scientists who are using the acoustic data for the study of marine
mammals~\cite{bib:nature_whales,bib:die_Zeit} . 
For the acoustic detection of
neutrinos, this cooperation helps to understand and reduce the
background from marine mammals.
As the AMADEUS device provides a constant stream of acoustic data to
shore---currently a unique feature in the Mediterranean Sea---marine
scientists 
are already using the AMADEUS data for dedicated research.
A live stream of the data from an AMA\-DE\-US hydrophone combined with a real-time
analysis focused on marine science
is available on the Internet~\cite{bib:lido-web}.

\section{Acoustic Sensing for KM3NeT}
\label{sec:km3net}

%
%
%
%
%
%
%
%

\label{sec:acoustic}
The planned towers\footnote{The vertical structures of the KM3NeT detector holding optical sensors are
called towers.} of the KM3NeT detector will be free to sway and 
twist in the undersea current with an expected displacement of up to
150\,m at the top for sea currents reaching 30\,cm per
second~\cite{bib:kooijman-km3net-icrc2011}.  In order to determine the
relative positions of the storeys with a precision of not worse than
20\,cm the detector will be equipped with an acoustic positioning
system.  The system employs acoustic transceivers on the sea floor and
acoustic receivers (hydrophones) in each storey.  By performing
multiple time-delay measurements and using these to triangulate the
positions of the individual hydrophones, the hydrophone positions can
be reconstructed relative to the positions of the emitters.  
%
Reliable, low-power and cost-efficient emitters operating in the 
range $20-40\,\mathrm{kHz}$ are under development~\cite{bib:ardid-vlvnt09}.

The KM3NeT positioning system is based on experience of the systems
developed for ANTARES, 
see~\cite{bib:antares-pos,bib:ameli-vlvnt09} and references therein.
Sampling will be done at about 200\,k samples per second and all data
will be written to shore. This way, algorithms for the position
calibration, running on an on-shore computer farm, can be adapted to
in-situ conditions that may affect the shape of the received signal.
Furthermore, the data can be used for additional analyses, such as
acoustic detection of neutrinos, see
e.g.~\cite{bib:nahnhauer-arena2010} and references therein, or marine
science investigations.

\begin{figure}
\begin{centering}
\subfigure[]{
\includegraphics[height=28.0mm]{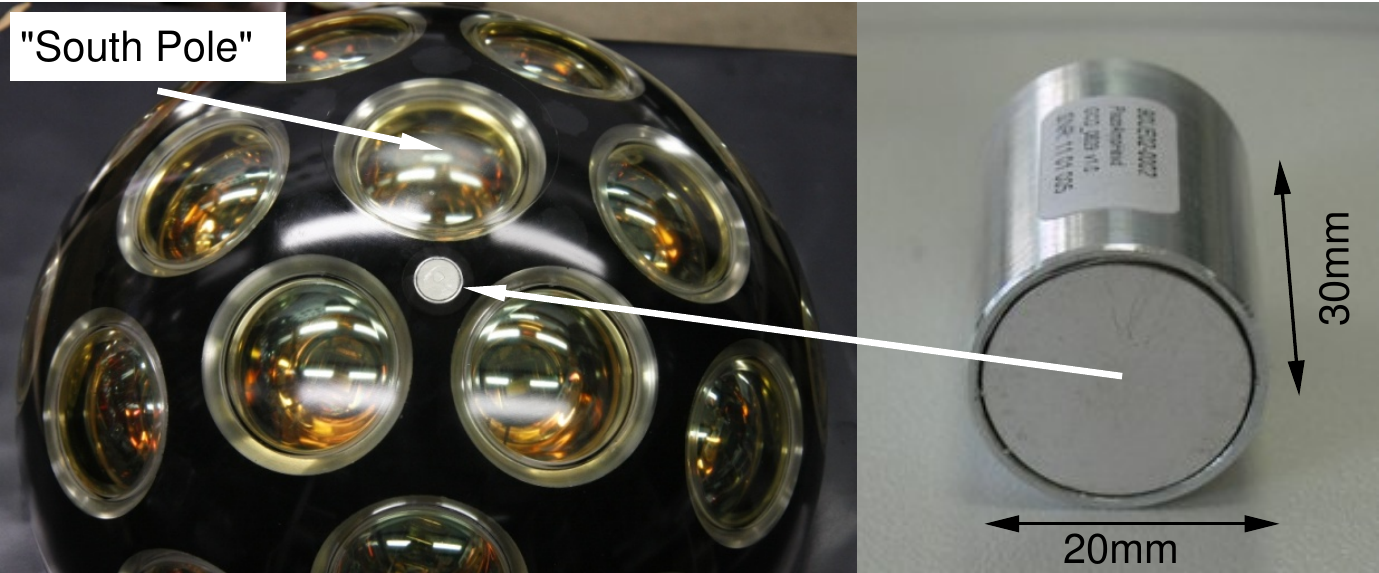}
\label{subfig:piezo_in_dome}
}
\subfigure[]{
\includegraphics[height=28.0mm,angle=0]{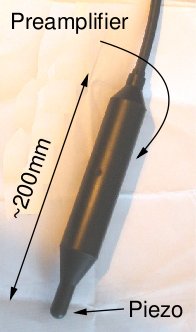}
\label{subfig:hydro_for_km3net}
}
\caption{
\subref{subfig:piezo_in_dome} Acoustic piezo sensor for installation
inside a DOM and a DOM with installed sensor.
\subref{subfig:hydro_for_km3net} Prototype of a hydrophone for KM3NeT,
to be mounted on the bar structure near the DOM.
}
\label{fig:positioning}
\end{centering}
\end{figure}

Two types of acoustic sensors, both based on the piezo-electric
effect, are currently tested. 
They are shown in Fig.~\ref{fig:positioning}.
The first kind are standard hydrophones, i.e.\ a
piezo ceramic and a preamplifier coated with polyurethane for high
pressure water tightness.  The second kind are compact units of a piezo ceramic and a
preamplifier, glued to the inside of the glass sphere of the DOM near
its ``South Pole''.  This design has been tested within
AMADEUS. 
The advantages w.r.t.\ to standard hydrophones are lower costs and a
reduction of the number of failure points: no additional cables and
junctions are required and the sensor is not exposed to the aggressive
environmental conditions.  Disadvantageous on the other hand is a
reduced angular acceptance and the vulnerability of the system to
electric interferences with the PMTs in the same sphere.  A 
KM3NeT prototype
Detection Unit planned to be deployed in 2012 will contain both types
of sensors and will allow for a design decision, 
see also~\cite{bib:enzenhoefer-vlvnt2011}.

\section{Conclusions and Outlook}
\label{sec:summary}

Acoustic detection is a promising approach for a future large volume 
detector of UHE neutrinos. To investigate the feasibility and potential of such a 
detector, several experiments have been performed or are underway.
These experiments use either existing military acoustic arrays or are additions to 
Cherenkov neutrino telescopes. Their sizes are far too small to yield competitive limits on 
the flux of UHE neutrinos but they allow for the investigation of experimental techniques
for a future acoustic neutrino detector and for the investigation of background conditions, which are the
essential factor that determines the feasibility of such a device.
At the same time, the continous stream of data from the deep sea provided by 
deep-sea acoustic arrays
is of great interest for marine scientists for the study of sea mammals.  

Background conditions in the Mediterranean Sea have been monitored by the AMADEUS
system within the ANTARES neutrino telescope. 
The ambient background was found to be stable at the expected level.
The transient background is very diverse and stems mainly from dolphins and 
shipping traffic. Methods for its suppression are under development, as are
Monte Carlo simulations and algorithms for neutrino selection.

For the proposed
KM3NeT neutrino telescope, a combined system for acoustic position calibration of the photomultipliers 
and neutrino
detection is planned. For the latter purpose, it would be an intermediate step towards
an even bigger acoustic detector for UHE neutrinos. 

\section{Acknowledgements}
The author wishes to thank the organizers of the VLVnT workshop for
the invitation to give a presentation. The AMADE\-US project is supported
by the German government (Bundesministerium f\"ur Bildung und
Forschung, BMBF) through grants 05A08WE1 and 05A11WE1.





\bibliographystyle{model1a-num-names}
\bibliography{vlvnt2011_lahmann}

\begin{thebibliography}{46}
\expandafter\ifx\csname natexlab\endcsname\relax\def\natexlab#1{#1}\fi
\providecommand{\bibinfo}[2]{#2}
\ifx\xfnm\relax \def\xfnm[#1]{\unskip,\space#1}\fi
\bibitem[{{R.J.\ Urick}(1983)}]{urick}
\bibinfo{author}{{R.J.\ Urick}}, \bibinfo{title}{Principles of Underwater
  Sound}, \bibinfo{publisher}{Peninsula publishing}, \bibinfo{year}{1983}.
  \bibinfo{note}{{ISBN 0-932146-62-7}}.
\bibitem[{{G.~Manthei, J.~Eisenbl\"{a}tter, and
  T.~Spies}(2006)}]{bib:salt-acoustics-arena2005}
\bibinfo{author}{{G.~Manthei, J.~Eisenbl\"{a}tter, and T.~Spies}}, in:
  \bibinfo{booktitle}{Proceedings of the 1st International Workshop on Acoustic
  and Radio EeV Neutrino detection Activities (ARENA 2005), Zeuthen, Germany,
  May 17--19}, Int. J. Mod. Phys. A21S1, \bibinfo{publisher}{World Scientific},
  \bibinfo{year}{2006}, p.~\bibinfo{pages}{30}. \bibinfo{note}{{ISBN
  981-256-755-0}}.
\bibitem[{{P.B. Price}(2006)}]{bib:price-2006}
\bibinfo{author}{{P.B. Price}}, \bibinfo{journal}{J.\ of Geophys.\ Res}
  \bibinfo{volume}{111} (\bibinfo{year}{2006}) \bibinfo{pages}{B02201}.
  \bibinfo{note}{{arXiv:astro-ph/0506648v1}}.
\bibitem[{{R.~Nahnhauer, A.A.~Rostovtsev, and
  D.~Tosi}(2008)}]{bib-permafrost-2008}
\bibinfo{author}{{R.~Nahnhauer, A.A.~Rostovtsev, and D.~Tosi}},
  \bibinfo{journal}{Nucl.\ Inst.\ and Meth.} \bibinfo{volume}{A\,587}
  (\bibinfo{year}{2008}) \bibinfo{pages}{29}. \bibinfo{note}{{arXiv:0707.3757v1
  [astro-ph]}}.
\bibitem[{{G.A.\ Askariyan, B.A.\ Dolgoshein, A.N.~Kalinovsky, and N.V.\
  Mokhov}(1979)}]{Askariyan2}
\bibinfo{author}{{G.A.\ Askariyan, B.A.\ Dolgoshein, A.N.~Kalinovsky, and N.V.\
  Mokhov}}, \bibinfo{journal}{Nucl.\ Inst.\ and Meth.} \bibinfo{volume}{164}
  (\bibinfo{year}{1979}) \bibinfo{pages}{267}.
\bibitem[{{J.G.~Learned}(1979)}]{Learned}
\bibinfo{author}{{J.G.~Learned}}, \bibinfo{journal}{Phys.\ Rev.}
  \bibinfo{volume}{D\,19} (\bibinfo{year}{1979}) \bibinfo{pages}{3293}.
\bibitem[{{S.~Bevan \etal\ (ACoRNE Coll.)}(2007)}]{bib:Sim_Acorne}
\bibinfo{author}{{S.~Bevan \etal\ (ACoRNE Coll.)}},
  \bibinfo{journal}{Astropart.\ Phys.} \bibinfo{volume}{28}
  (\bibinfo{year}{2007}) \bibinfo{pages}{366}.
  \bibinfo{note}{{arXiv:0704.1025v1 [astro-ph]}}.
\bibitem[{{S.~Bevan \etal\ (ACoRNE Coll.)}(2009)}]{bib:Sim_Acorne2}
\bibinfo{author}{{S.~Bevan \etal\ (ACoRNE Coll.)}}, \bibinfo{journal}{Nucl.\
  Inst.\ and Meth.} \bibinfo{volume}{A 607} (\bibinfo{year}{2009})
  \bibinfo{pages}{398}. \bibinfo{note}{{arXiv:0903.0949v2 [astro-ph.IM]}}.
\bibitem[{{V.\ Niess and V.\ Bertin}(2006)}]{bib:Bertin_Niess}
\bibinfo{author}{{V.\ Niess and V.\ Bertin}}, \bibinfo{journal}{Astropart.\
  Phys.} \bibinfo{volume}{26} (\bibinfo{year}{2006}) \bibinfo{pages}{243}.
  \bibinfo{note}{{arXiv:astro-ph/0511617v3}}.
\bibitem[{{A.\ Roberts}(1992)}]{bib:roberts-1992}
\bibinfo{author}{{A.\ Roberts}}, \bibinfo{journal}{Rev.\ Mod.\ Phys.}
  \bibinfo{volume}{64} (\bibinfo{year}{1992}) \bibinfo{pages}{259}.
\bibitem[{{L.G.\ Dedenko \etal}(1997)}]{bib:sadco-1997}
\bibinfo{author}{{L.G.\ Dedenko \etal}}, \bibinfo{title}{{Sadco: Hydroacoustic
  Detection of Super-High Energy Cosmic Neutrinos}}, \bibinfo{year}{1997}.
  \bibinfo{note}{{arXiv:astro-ph/9705189v1}}.
\bibitem[{{R.~Nahnhauer}(2012)}]{bib:nahnhauer-arena2010}
\bibinfo{author}{{R.~Nahnhauer}}, in: \bibinfo{booktitle}{{Proceedings of the
  4th International Workshop on Acoustic and Radio EeV Neutrino Detection
  Activities (ARENA 2010), Nantes, France, June 29--July 2}}, Nucl.\ Inst.\ and
  Meth.\ A\,662, p. \bibinfo{pages}{S20}. \bibinfo{note}{{arXiv:1010.3082v2
  [astro-ph.IM]}}.
\bibitem[{{A.~Achterberg~\etal\ (IceCube
  Coll.)}(2006)}]{bib:icecube-app:26:155}
\bibinfo{author}{{A.~Achterberg~\etal\ (IceCube Coll.)}},
  \bibinfo{journal}{Astropart.\ Phys.} \bibinfo{volume}{26}
  (\bibinfo{year}{2006}) \bibinfo{pages}{155}.
  \bibinfo{note}{{arXiv:astro-ph/0604450}}.
\bibitem[{{A.~Avrorin~\etal}(2011)}]{bib:baikal-nt-vlvnt09}
\bibinfo{author}{{A.~Avrorin~\etal}}, in: \bibinfo{booktitle}{{Proceedings of
  the 4th International Workshop on a Very Large Volume Neutrino Telescope for
  the Mediterranean Sea (VLVnT 2009), Athens, Greece, 13--15 Oct. 2009}},
  Nucl.\ Inst.\ and Meth.\ A\,626--627, p. \bibinfo{pages}{S13}.
\bibitem[{{M.~Taiuti~\etal}(2011)}]{bib:nemo-vlvnt09}
\bibinfo{author}{{M.~Taiuti~\etal}}, in: \bibinfo{booktitle}{{Proceedings of
  the 4th International Workshop on a Very Large Volume Neutrino Telescope for
  the Mediterranean Sea (VLVnT 2009), Athens, Greece, 13--15 Oct. 2009}},
  Nucl.\ Inst.\ and Meth.\ A\,626--627, p. \bibinfo{pages}{S25}.
\bibitem[{{M. Ageron~\etal\ (ANTARES Coll.)}(2011)}]{bib:ANTARES-paper}
\bibinfo{author}{{M. Ageron~\etal\ (ANTARES Coll.)}}, \bibinfo{journal}{Nucl.\
  Inst.\ and Meth.} \bibinfo{volume}{A 656} (\bibinfo{year}{2011})
  \bibinfo{pages}{11}. \bibinfo{note}{{arXiv:1104.1607v1 [astro-ph.IM]}}.
\bibitem[{{T.~Karg, for the IceCube Coll.}(2012)}]{bib:karg-arena2010}
\bibinfo{author}{{T.~Karg, for the IceCube Coll.}}, in:
  \bibinfo{booktitle}{{Proceedings of the 4th International Workshop on
  Acoustic and Radio EeV Neutrino Detection Activities (ARENA 2010), Nantes,
  France, June 29-July 2}}, Nucl.\ Inst.\ and Meth.\ A\,662, p.
  \bibinfo{pages}{S36}. \bibinfo{note}{{arXiv:1010.2025v1 [astro-ph.IM]}}.
\bibitem[{{Y.~Abdou~\etal\ (IceCube
  Coll.)}(2012)}]{bib:SPATS-design-perform-2011}
\bibinfo{author}{{Y.~Abdou~\etal\ (IceCube Coll.)}}, \bibinfo{journal}{Nucl.\
  Inst.\ and Meth.\ A} \bibinfo{volume}{683} (\bibinfo{year}{2012})
  \bibinfo{pages}{78}. \bibinfo{note}{{arXiv:1105.4339v1 [astro-ph.IM]}}.
\bibitem[{{R.~Abbasi \etal\ (IceCube Coll.)}(2010)}]{bib:spats-2010-speed}
\bibinfo{author}{{R.~Abbasi \etal\ (IceCube Coll.)}},
  \bibinfo{journal}{Astropart.\ Phys.} \bibinfo{volume}{33}
  (\bibinfo{year}{2010}) \bibinfo{pages}{277}.
  \bibinfo{note}{{arXiv:0909.2629v1 [astro-ph.IM]}}.
\bibitem[{{R.~Abbasi \etal\ (IceCube Coll.)}(2011)}]{bib:spats-2011-atten}
\bibinfo{author}{{R.~Abbasi \etal\ (IceCube Coll.)}},
  \bibinfo{journal}{Astropart.\ Phys.} \bibinfo{volume}{34}
  (\bibinfo{year}{2011}) \bibinfo{pages}{382}.
  \bibinfo{note}{{arXiv:1004.1694v2 [astro-ph.IM]}}.
\bibitem[{{R.~Abbasi \etal\ (IceCube Coll.)}(2012)}]{bib:spats-2011-bkgr}
\bibinfo{author}{{R.~Abbasi \etal\ (IceCube Coll.)}},
  \bibinfo{journal}{Astropart.\ Phys.} \bibinfo{volume}{35}
  (\bibinfo{year}{2012}) \bibinfo{pages}{312}.
  \bibinfo{note}{{arXiv:1103.1216v1 [astro-ph.IM]}}.
\bibitem[{{K.~Antipin \etal\ (BAIKAL Coll.)}(2007)}]{bib:baikal}
\bibinfo{author}{{K.~Antipin \etal\ (BAIKAL Coll.)}}, in:
  \bibinfo{booktitle}{Proceedings of the 30th International Cosmic Ray
  Conference (ICRC2007), Merida, Mexico, July 3--11}.
  \bibinfo{note}{{arXiv:0710.3113 [astro-ph]}}.
\bibitem[{{V.~Aynutdinov~\etal}(2010)}]{bib:baikal-arena2010}
\bibinfo{author}{{V.~Aynutdinov~\etal}}, in: \bibinfo{booktitle}{{Proceedings
  of the 4th International Workshop on Acoustic and Radio EeV Neutrino
  Detection Activities (ARENA 2010), Nantes, France, June 29--July 2}}, Nucl.\
  Inst.\ and Meth.\ A. \bibinfo{note}{Doi:10.1016/j.nima.2010.11.153}.
\bibitem[{{V.~Aynutdinov~\etal}(2009)}]{bib:baikal-acoust-icrc09}
\bibinfo{author}{{V.~Aynutdinov~\etal}}, in: \bibinfo{booktitle}{Proceedings of
  the 31st Int. Cosmic Ray Conf. (ICRC2009), Lodz, Poland, July 7--15}.
  \bibinfo{note}{{arXiv:0910.0678v1 [astro-ph.HE]}}.
\bibitem[{{G.\ Riccobene for the NEMO Coll.}(2009)}]{bib:noise_ONDE}
\bibinfo{author}{{G.\ Riccobene for the NEMO Coll.}}, in:
  \bibinfo{booktitle}{{Proceedings of the 3rd International Workshop on
  Acoustic and Radio EeV Neutrino Detection Activities (ARENA 2008), Rome,
  Italy, June 25--27}}, Nucl.\ Inst.\ and Meth.\ A\,604, p.
  \bibinfo{pages}{149}.
\bibitem[{{P.W.~Gorham~\etal\ (ANITA Coll.)}(2010)}]{bib:anita-2010}
\bibinfo{author}{{P.W.~Gorham~\etal\ (ANITA Coll.)}}, \bibinfo{journal}{Phys.\
  Rev.} \bibinfo{volume}{D\,82} (\bibinfo{year}{2010}) \bibinfo{pages}{022004}.
  \bibinfo{note}{{arXiv:1003.2961v3 [astro-ph.HE]; Erratum: arXiv:1011.5004v1
  [astro-ph.HE]}}.
\bibitem[{{N.G.~Lehtinen, P.W.~Gorham, A.R.~Jacobson, and
  R.A.~Roussel-Dupr\'e}(2004)}]{bib:forte-2004}
\bibinfo{author}{{N.G.~Lehtinen, P.W.~Gorham, A.R.~Jacobson, and
  R.A.~Roussel-Dupr\'e}}, \bibinfo{journal}{Phys.\ Rev.}
  \bibinfo{volume}{D\,69} (\bibinfo{year}{2004}) \bibinfo{pages}{013008}.
  \bibinfo{note}{{arXiv:astro-ph/0309656v2}}.
\bibitem[{{P.W.~Gorham~\etal}(2004)}]{bib:glue-2004}
\bibinfo{author}{{P.W.~Gorham~\etal}}, \bibinfo{journal}{Phys.\ Rev.\ Lett.}
  \bibinfo{volume}{93} (\bibinfo{year}{2004}) \bibinfo{pages}{041101}.
  \bibinfo{note}{{arXiv:astro-ph/0310232v3}}.
\bibitem[{{N.~Kurahashi, J.~Vandenbroucke, and
  G.~Gratta}(2010)}]{bib:saund2010}
\bibinfo{author}{{N.~Kurahashi, J.~Vandenbroucke, and G.~Gratta}},
  \bibinfo{journal}{Phys.\ Rev.} \bibinfo{volume}{D\,82} (\bibinfo{year}{2010})
  \bibinfo{pages}{073006}. \bibinfo{note}{{arXiv:1007.5517v1 [hep-ex]}}.
\bibitem[{{S.~Bevan}(2009)}]{bib:bevan_ARENA08}
\bibinfo{author}{{S.~Bevan}}, in: \bibinfo{booktitle}{{Proceedings of the 3rd
  International Workshop on Acoustic and Radio EeV Neutrino Detection
  Activities (ARENA 2008), Rome, Italy, June 25--27}}, Nucl.\ Inst.\ and Meth.\
  A\,604, p. \bibinfo{pages}{143}.
\bibitem[{{R. Engel, D. Seckel, and T. Stanev}(2001)}]{bib:EES-GZKnus-2001}
\bibinfo{author}{{R. Engel, D. Seckel, and T. Stanev}},
  \bibinfo{journal}{Phys.\ Rev.} \bibinfo{volume}{D\,64} (\bibinfo{year}{2001})
  \bibinfo{pages}{093010}. \bibinfo{note}{{arXiv:astro-ph/0101216v2}}.
\bibitem[{{D. Seckel}(2005)}]{bib:seckel-gzk-www-reduced}
\bibinfo{author}{{D. Seckel}}, \bibinfo{howpublished}{Internet},
  \bibinfo{year}{2005}. \bibinfo{note}{{\it
  ftp://ftp.bartol.udel.edu/../seckel/ess-gzk/}; ``Proton model, Auger'': file
  2008/export\_data/ss\_135\_a.txt; ``Proton model, HiRes'': file
  2008/export\_data/ss\_135\_h.txt; ``Mixed comp model, HiRes'': file
  2008/export\_data/ss\_151\_h.txt}.
\bibitem[{{N.G. Lehtinen \etal}(2002)}]{bib:Lehtinen-2002}
\bibinfo{author}{{N.G. Lehtinen \etal}}, \bibinfo{journal}{Astropart.\ Phys.}
  \bibinfo{volume}{17} (\bibinfo{year}{2002}) \bibinfo{pages}{279}.
  \bibinfo{note}{{arXiv:astro-ph/0104033v1}}.
\bibitem[{{J. Vandenbroucke, G. Gratta, and N. Lehtinen}(2005)}]{bib:saund2004}
\bibinfo{author}{{J. Vandenbroucke, G. Gratta, and N. Lehtinen}},
  \bibinfo{journal}{Astrophys.\ J.} \bibinfo{volume}{621}
  (\bibinfo{year}{2005}) \bibinfo{pages}{301}.
  \bibinfo{note}{{arXiv:astro-ph/0406105v2}}.
\bibitem[{{S.~Danaher for the ACoRNE Coll. }(2007)}]{bib:acorne}
\bibinfo{author}{{S.~Danaher for the ACoRNE Coll. }}, in:
  \bibinfo{booktitle}{Proceedings of the 2nd International Workshop on Acoustic
  and Radio EeV Neutrino detection Activities (ARENA 2006), Newcastle, UK,
  28--30 June}, volume~\bibinfo{volume}{81} of \textit{\bibinfo{series}{J.\
  Phys.\ Conf.\ Ser.}}, \bibinfo{publisher}{IOP Publishing, Philadelphia},
  \bibinfo{year}{2007}, p. \bibinfo{pages}{012011}.
\bibitem[{{J.A. Aguilar~\etal\ (ANTARES Coll.)}(2011)}]{bib:amadeus-2010}
\bibinfo{author}{{J.A. Aguilar~\etal\ (ANTARES Coll.)}},
  \bibinfo{journal}{Nucl.\ Inst.\ and Meth.} \bibinfo{volume}{A 626-627}
  (\bibinfo{year}{2011}) \bibinfo{pages}{128}.
  \bibinfo{note}{{arXiv:1009.4179v2 [astro-ph.IM]}}.
\bibitem[{{M.~Neff \etal}(2010)}]{bib:neff_arena2010}
\bibinfo{author}{{M.~Neff \etal}}, in: \bibinfo{booktitle}{{Proceedings of the
  4th International Workshop on Acoustic and Radio EeV Neutrino Detection
  Activities (ARENA 2010), Nantes, France, June 29--July 2}}, Nucl.\ Inst.\ and
  Meth.\ A\,662, p. \bibinfo{pages}{S242}. \bibinfo{note}{{arXiv:1104.3248v1
  [astro-ph.IM]}}.
\bibitem[{{M. Neff}(shed)}]{bib:neff-vlvnt2011}
\bibinfo{author}{{M. Neff}}, in: \bibinfo{booktitle}{{Proceedings of the 5th
  International Workshop on a Very Large Volume Neutrino Telescope for the
  Mediterranean Sea (VLVnT 2011), Erlangen, Germany, 12--14 Oct. 2011}}, Nucl.\
  Inst.\ and Meth.\ A, to be published.
\bibitem[{Nosengo(2009)}]{bib:nature_whales}
\bibinfo{author}{N.~Nosengo}, \bibinfo{journal}{Nature} \bibinfo{volume}{462}
  (\bibinfo{year}{2009}) \bibinfo{pages}{560}. \bibinfo{note}{{News Feature}}.
\bibitem[{{H.~Rietz}(2011)}]{bib:die_Zeit}
\bibinfo{author}{{H.~Rietz}}, \bibinfo{title}{{Walgesang als Beifang}},
  \bibinfo{howpublished}{published in the German newspaper ``Die Zeit'' on
  March 14th}, \bibinfo{year}{{2011}}. \bibinfo{note}{{\it
  http://www.zeit.de/2011/11/N-Neutrinodetektoren}}.
\bibitem[{{Web page of the LIDO (Listening to the Deep-Ocean Environment)
  project}(2011)}]{bib:lido-web}
\bibinfo{author}{{Web page of the LIDO (Listening to the Deep-Ocean
  Environment) project}}, \bibinfo{year}{2011}. \bibinfo{note}{{\it
  http://listentothedeep.org/}}.
\bibitem[{{P. Kooijman for the KM3NeT
  Consortium}(2011)}]{bib:kooijman-km3net-icrc2011}
\bibinfo{author}{{P. Kooijman for the KM3NeT Consortium}}, in:
  \bibinfo{booktitle}{{Proceedings of the 32nd Int. Cosmic Ray Conf.
  (ICRC2011), Beijing, China, Aug. 11--18}}.
\bibitem[{{M.~Ardid~\etal\ for the KM3NeT
  Consortium}(2011)}]{bib:ardid-vlvnt09}
\bibinfo{author}{{M.~Ardid~\etal\ for the KM3NeT Consortium}}, in:
  \bibinfo{booktitle}{{Proceedings of the 4th International Workshop on a Very
  Large Volume Neutrino Telescope for the Mediterranean Sea (VLVnT 2009),
  Athens, Greece, 13--15 Oct. 2009}}, Nucl.\ Inst.\ and Meth.\ A\,626--627, p.
  \bibinfo{pages}{S214}.
\bibitem[{{M.\ Ardid for the ANTARES Coll.}(2009)}]{bib:antares-pos}
\bibinfo{author}{{M.\ Ardid for the ANTARES Coll.}}, in:
  \bibinfo{booktitle}{Proceedings of the 3rd International Workshop on a Very
  Large Volume Neutrino Telescope for the Mediterranean Sea (VLVnT 2008),
  Toulon, France, April 22--24}, Nucl.\ Inst.\ and Meth.\ A\,602, p.
  \bibinfo{pages}{174}.
\bibitem[{{F.~Ameli~\etal\ for the KM3NeT
  Consortium}(2011)}]{bib:ameli-vlvnt09}
\bibinfo{author}{{F.~Ameli~\etal\ for the KM3NeT Consortium}}, in:
  \bibinfo{booktitle}{{Proceedings of the 4th International Workshop on a Very
  Large Volume Neutrino Telescope for the Mediterranean Sea (VLVnT 2009),
  Athens, Greece, 13--15 Oct. 2009}}, Nucl.\ Inst.\ and Meth.\ A\,626--627, p.
  \bibinfo{pages}{S211}.
\bibitem[{{A. Enzenh\"ofer}(shed)}]{bib:enzenhoefer-vlvnt2011}
\bibinfo{author}{{A. Enzenh\"ofer}}, in: \bibinfo{booktitle}{{Proceedings of
  the 5th International Workshop on a Very Large Volume Neutrino Telescope for
  the Mediterranean Sea (VLVnT 2011), Erlangen, Germany, 12--14 Oct. 2011}},
  Nucl.\ Inst.\ and Meth.\ A, to be published.

\end{thebibliography}







\end{document}